\documentclass[12pt]{article}
\parskip=\medskipamount

\usepackage{hyperref}
\usepackage{amssymb}
\usepackage{graphicx}

\def\caja{\mathsurround=0pt}
\def\eqalign#1{\,\vcenter{\openup2\jot \caja
        \ialign{\strut \hfil$\displaystyle{##}$&$
        \displaystyle{{}##}$\hfil\crcr#1\crcr}}\,}
\newif\ifdtup

\def\Bar#1{\overline{#1}}

\newcommand {\cA}{{\cal A}}

\newcommand {\cC}{{\cal C}}

\newcommand {\cN}{{\cal N}}

\newcommand {\cV}{{\cal V}}

\def\a{\alpha}
\def \bi{\bibitem}

\def\b{\beta}

\def\d{\delta}

\def\g{\gamma}
\def\G{\Gamma}

\def\p{\pi}

\def\z{\zeta}
\def\D{\Delta}
\def\F{\Phi}
\def\J{\Psi}
\def\L{\Lambda}

\def\S{\Sigma}
\def\U{\Upsilon}

\newcommand{\ad}{{\dot{\alpha}}}                           
\newcommand{\bd}{{\dot{\beta}}}                            
\newcommand{\ve}{\varepsilon}                            
\newcommand{\gd}{{\dot{\gamma}}}

\newcommand{\pa}{\partial}                           
\newcommand{\hf}{\frac12}

\newcommand{\vf}{\varphi}
\newcommand{\sect}[1]{\setcounter{equation}{0}\section{#1}}

\newcommand{\be}{\begin{equation}}
\newcommand{\ee}{\end{equation}}
\newcommand{\bea}{\begin{eqnarray}}
\newcommand{\eea}{\end{eqnarray}}
\newcommand{\non}{\nonumber}
\newcommand{\1}{\underline{1}}

\def \intss{\int\!\!{\rm d}^8z}

\newcommand{\bm}[1]{\mbox{\boldmath$#1$}}

\begin{document}

\begin{titlepage}

\begin{flushright}
hep-th/0506255\\
UMDEPP 05-046\\
June, 2005
\end{flushright}
\vspace{5mm}

\begin{center}
{\Large\bf  
$\bm 4$D, $\bm {{\cal N} \,=\, 1}$
Higher Spin Gauge Superfields\\
and Quantized Twistors
}
\end{center}

\begin{center}

{\large 
S. James Gates, Jr.$^\dag$\footnote{gatess@wam.umd.edu}
and Sergei M. 
Kuzenko\footnote{kuzenko@cyllene.uwa.edu.au}${}^\ddagger$
}
\vspace{2mm}

${}^\dag$\footnotesize{
{\it Center for String and Particle Theory\\
Department of Physics, University of Maryland\\
College Park, MD 20742-4111 USA}}\\
~\\
${}^\ddagger$\footnotesize{
{\it School of Physics M013, The University of Western Australia,\\
35 Stirling Highway, Crawley W.A. 6009, Australia}} 
\vspace{2mm}

\end{center}
\vspace{5mm}

\begin{abstract}
\baselineskip=14pt
${}$For the gauge massless higher spin 4D, $\cal N$ = 1 
off-shell supermultiplets previously developed, 
we provide evidence of a twistor-like oscillator realization 
that  is intrinsically related 
to the superfield structure of the dynamical variables 
and gauge transformations. 
Gauge invariant field strengths and 
linearized Bianchi identities for these multiplets 
are worked out. It is further argued, inspired 
by earlier non-supersymmetric constructions due to 
Klishevich and Zinoviev,  that a massive 
superspin-$s$ multiplet can  be described as 
a gauge-invariant dynamical system involving 
massless  multiplets of superspins $s,s-1/2, \ldots ,0$.  
A gauge-invariant formulation for the massive gravitino 
multiplet is discussed in some detail.
\end{abstract}

\vfill
\end{titlepage}

\newpage
\setcounter{page}{1}
\renewcommand{\thefootnote}{\arabic{footnote}}
\setcounter{footnote}{0}

\sect{Introduction}

$~~~~$ In four space-time dimensions,
Lagrangian formulations for massive fields
of arbitrary  spin were constructed thirty years
ago \cite{SH}, as a partial realization of the Fierz-Pauli 
program \cite{FP}.  A few years later, the Singh-Hagen models 
\cite{SH} were used to derive
Lagrangian formulations for gauge massless fields
of arbitrary spin \cite{Fronsdal1}. 
The massless construction of \cite{Fronsdal1} 
was then extended to (anti) de Sitter space 
\cite{Fronsdal2}, as well
it stimulated the appearance of elegant
reformulations  and generalizations, 
see e. g. \cite{deWF,Vas1}. 

In supersymmetric field theory, 
the supersymmetric analogue
of  the Casimir operator spin 
is called the {\it superspin} \cite{Likhtman,SS,Sok}
(similarly, there exists natural supersymmetric extensions 
of the helicity \cite{GGRS,BK,PZ}).
It is therefore of some interest to develop supersymmetric 
extensions of the models discovered in \cite{SH,Fronsdal1}.
As was first demonstrated in the work of \cite{Curtright} 
and later in  \cite{Vas1},
on-shell massless multiplets of arbitrary superspin are 
easily obtained by  putting together two massless 
spin-$s$ and spin-$(s+1/2)$ actions,  derived 
in \cite{Fronsdal1} or \cite{Vas1}, 
and  then guessing the structure of supersymmetry transformations.
In such an approach, however, 
the supersymmetry transformations form a closed algebra
only on the mass shell. It proves  to be more difficult 
to construct off-shell massless higher superspin multiplets.
The latter problem was solved in \cite{KSP,KS} (see \cite{BK}
for a review) building  on the prepotential structure 
of $\cN=1$ superfield supergravity \cite{SG} 
(see \cite{GGRS,BK} for reviews) as a guiding principle. 
${}$For each superspin $s>3/2$, these
publications provide two dually equivalent off-shell
realizations in  4D, ${\cal N }= 1$ superspace.
At the component level, each of the
two superspin-$s$ actions \cite{KSP,KS} reduces, 
{\it upon} imposing a Wess-Zumino-type
gauge and eliminating the auxiliary fields,
to a sum of the spin-$s$
and  spin-$(s+1/2)$ actions \cite{Fronsdal1}.
On the mass shell, the only independent
gauge-invariant field strengths in these models
are  exactly the higher spin on-shell field strengths
first identified in ``Superspace''  \cite{GGRS}.
The gauge massless  higher spin  supermultiplets of \cite{KSP,KS}
were also  generalized to $\cN=1$ anti-de Sitter superspace \cite{KS2}. 
In addition,  there have been developed 
4D, $\cN=2$ {\it off-shell} massless higher spin supermultiplets
\cite{GKS} (see also \cite{SeSi}),  
as well as a generating superfield action for arbitrary superspin
massless multiplets in 
4D, $\cN=1$ anti-de Sitter superspace \cite{GKS2}.

In the massive case,  higher spin
supermultiplets have never been constructed 
beyond superspin-3/2\footnote{As is well known
\cite{Likhtman,SS}, 
a massive $\cN=1$ multiplet of superspin $s$
describes four propagating \newline $~~~~~~$ fields
with the same mass but different
spins $(s-1/2, s, s, s+1/2)$,
see \cite{GGRS,BK} for reviews.};
only the cases of 
massive gravitino multiplet 
(superspin-1) and massive graviton 
multiplet (superspin-3/2) have been studied 
in some detail
\cite{OS2,FvN,BL,AB,BGLP,Zinoviev2,GSS,BGKP,Ph}, 
both at the off-shell and on-shell levels.
${}$For constructing massive higher spin 
supermultiplets, one could try to develop 
a direct extension of the Singh-Hagen approach \cite{SH}. 
However it seems less formidable, 
and also conceptually very appealing, 
to look for a supersymmetric extension of the approach 
advocated a few years ago by Klishevich and Zinoviev 
\cite{KZ,Zinoviev}. In their approach, 
the massive spin-$s$ particle is described
by a gauge-invariant action which involves all the massless
fields  \cite{Fronsdal1} of spins $s, s-1, \dots$ ,
and possesses the properties: (i) in the massless limit, 
the action becomes a sum of the massless actions
\cite{Fronsdal1} of spins $s, s-1, \dots$; 
(ii) the gauge symmetry is a mass-dependent deformation 
of the massless gauge transformations; 
(iii) the gauge freedom can be used to choose a 
Wess-Zumino-type gauge condition in which 
the action reduces to the massive spin-$s$ action 
of \cite{SH}.
In a sense, the scheme developed in  \cite{KZ,Zinoviev} 
is a higher spin analogue of the St\"uckelberg construction.
In the supersymmetric case, a gauge-invariant 
realization for the massive superspin-$s$ multiplet
should involve massless multiplets of superspins
$s,s-1/2, \dots ,0$.

Assuming the existence of a gauge-invariant 
formulation for massive higher superspin multiplets, 
the gauge massless models introduced in \cite{KSP,KS}
should clearly form a natural starting point. 
But for each massless superspin $s >3/2$, there are two 
dually equivalent  realizations (there exist three off-shell 
realizations for the massless multiplets of superspin 
$s=0 $ and $1$, and four off-shell realizations
for the massless superspin-3/2 multiplet). 
It is not clear a priori which one should occur 
as a building block in the construction  of massive 
supermultiplets\footnote{Massless $\cN=2$ supermultiplets 
are easier to construct \cite{GKS,GKS2}
using the transverse formulation  \newline $~~~~~~$ 
for half-integer superspins \cite{KSP} 
and the longitudinal formulation  for integer
superspins.}
(probably, several dually equivalent formulations also exist 
for massive higher superspin mulptiplets, as in the case 
of the massive vector multiplet (superspin-$1/2$)
described in Appendix B).  We are not able 
to definitively answer this and similar questions 
currently. We believe that there still remain some 
important properties of the massless higher superspin 
multiplets which have to be studied beforehand.
In addition, gauge-invariant descriptions for the  
massive gravitino multiplet (superspin-1) and massive 
graviton multiplet (superspin-3/2) should be studied 
in detail (to the best  of our knowledge, the observation 
given in \cite{AB} regarding the massive gravitino multiplet 
is the only result available).

This paper is organized as follows. 
In section 2 we review, following \cite{KSP,KS}, 
the superfield structure of massless higher 
superspin multiplets and their gauge symmetries.
Section 3 is devoted to a twistor-like 
oscillator realization that is intrinsically related 
to the superfield structure of the dynamical variables 
and gauge transformations introduced. Using the
quantized twistor, four BRST-like operators
are defined. The compatibility of introducing the
twistor-like oscillator realization is also discussed for
the usual 4D, $\cal N$ $=$ 1 Abelian gauge theory.
In sections 4 and 5, the structure of 
gauge-invariant field strengths and corresponding 
Bianchi identities, which occur in the massless 
higher superspin models is reviewed.
In  appendix A we collect the gauge-invariant action 
for the massless higher superspin multiplets 
\cite{KSP,KS}.  Three dually equivalent gauge-invariant  
realizations for the massive  gravitino multiplet are 
discussed in appendix B.  Finally, a gauge-invariant 
formulation for the massive gravitino multiplet is given 
in appendix C.

\sect{Higher Spin Gauge Superfields}

$~~~~$ The off-shell gauge 
formulations  \cite{KSP,KS}
for higher superspin 
massless multiplets in 4D, $\cN=1$ 
Minkowski superspace\footnote{Our superspace 
notation and conventions correspond to \cite {BK}, 
in particular the flat superspace
covari-  \newline $~~~~~~$ ant derivatives are
$D_A = (\pa_a, D_\a,\bar D^\ad)$. 
Throughout
this paper we consider only Lorentz 
tensors  \newline $~~~~~~$ symmetric in their undotted indices and separately 
in their dotted ones.   
${}$For a tensor of type $(k,l)$ \newline $~~~~~~$  with $k$ undotted 
and $l$ dotted indices we use the shorthand notations
$ 
\Psi_{\a(k) \ad(l)}   \equiv \Psi_{\a_1 \ldots
\a_k\ad_1\ldots \ad_l} =$  \newline $~~~~~~$  $ \Psi_{(\a_1 \ldots \a_k)(\ad_1\ldots
\ad_l)}$.
Quite often
we assume that the upper or lower indices,
which are denoted by  \newline $~~~~~~$ one and the same letter,
should be symmetrized, for instance
$
\phi_{\a(k)} \psi_{\a(l)}
\equiv \phi_{(\a_1\ldots\a_k} \psi_{\a_{k+1}\ldots\a_{k+l})}$.
 \newline $~~~~~~$ 
Given two tensors of the same type, their contraction is denoted by
$f \cdot g \equiv f^{\a(k) \ad(l)} \,g_{\a(k) \ad(l)}
= $  \newline $~~~~~~$  $f^{\a_1 \ldots \a_k\ad_1\ldots \ad_l}\, g_{\a_1 \ldots
\a_k\ad_1\ldots \ad_l}$.
} 
involve so-called transverse and longitudinal 
linear superfields, both as dynamical variables and 
gauge parameters. A complex tensor superfield 
$\G_{\a(k) \ad(l)}$ subject to the constraint
\bea
{\bar D}^\bd \,\G_{\a(k) \bd \ad(l-1)} & = & 0 ~, \qquad l>0 ~, 
\label{transverse}
\eea
is said to be transverse linear. A longitudinal linear superfield
$G_{\a(k) \ad(l)}$  is defined to satisfy the constraint
\be
{\bar D}_{ (\bd} \,G_{\a(k) \ad_1 \dots \ad_l)}=0 ~.
\label{longit}
\ee
The above constraints imply that
$\G_{\a(k) \ad(l)}$ and $G_{\a(k) \ad(l)}$ 
are linear in the usual sense
\be
{\bar D}^2\,  \G_{\a(k) \ad(l)} = {\bar D}^2 \,G_{\a(k) \ad(l)} = 0 ~.
\label{5}
\ee
In the case $l=0$, constraint  (\ref{transverse})
should be  replaced by ${\bar D}^2 \G_{\a(k) } =0$. Constraint 
(\ref{longit}) for $l=0 $ simply means that $G_{\a(k)} $ is chiral, 
${\bar D}_{ \bd} \,G_{\a(k) }=0$.
The constraints (\ref{transverse}) and 
(\ref{longit}) can be solved in terms of
unconstrained potentials
$ \Phi_{\a(k)\ad(l+1)} $ and $ \Psi_{\a(k)\ad(l-1)} $
as follows:
\bea
 \G_{\a(k) \ad(l)}&=& \bar D^\bd 
\Phi_{\a(k)\,\bd \ad(l ) } ~,
\qquad
 G_{\a(k) \ad(l)} ={\bar D}_{( \ad_l }
 \Psi_{ \a(k) \, \ad_1 \cdots \ad_{l-1})} ~.
\label{7}
\eea

Two formulations for the massless multiplet of 
a half-integer superspin 
$s+1/2$ (with $s=1,2\ldots$) which were called 
in Ref. \cite{KSP} transverse and longitudinal, 
contain the following dynamical variables respectively:
\bea
\cV^\bot_{s+1/2}& = &\Big\{H_{\a(s)\ad(s)}~, ~
\G_{\a(s-1) \ad(s-1)}~,
~ \bar{\G}_{\a(s-1) \ad(s-1)} \Big\} ~,    \\
\label{10}
\cV^{\|}_{s+1/2} &=& 
\Big\{H_{\a(s)\ad(s)}~, ~
G_{\a(s-1) \ad(s-1)}~,
~ \bar{G}_{\a(s-1) \ad(s-1)} \Big\}
~.
\label{11}
\eea
Here $H_{\a(s) \ad (s)}$ is real, 
$\G_{\a (s-1) \ad (s-1)} $ transverse linear and 
$G_{\a (s-1) \ad (s-1)}$ 
longitudinal linear  superfields.  
The  case $s=1$  corresponds to 
linearized supergravity  
(see  \cite{GGRS,BK} for  reviews).  

The gauge transformations for the superfields 
$H_{\a(s) \ad (s)} $, $\G_{\a (s-1) \ad(s-1)}$
and $G_{\a (s-1) \ad (s-1)}$  
postulated in \cite{KSP}
are 
\bea
\d  H_{\a(s) \ad (s)}  &=& g_{\a(s) \ad (s)}  
+ {\bar g}_{\a(s) \ad (s)}  ~,  
\label{16}\\
\d  \G_{\a (s-1) \ad(s-1)}
&=& \hf \, \frac{s}{s+1} \, \bar D^\bd D^\b 
{\bar g}_{\b \a(s-1) \bd \ad(s-1)}~,
\label{17}\\
\d G_{\a (s-1) \ad (s-1)}
&=& \hf \, \frac s{s+1} \, 
D^\b \bar D^\bd 
g _{\b \a(s-1) \bd \ad(s-1)} + {\rm i} \,s\,
 \pa^{\b \bd } g _{\b \a(s-1) \bd \ad(s-1)} ~, 
\label{18}
\eea
with a longitudinal linear parameter $g_{\a(s) \ad (s)}  $.
It can be seen that $\d G_{\a (s-1) \ad (s-1)}$ is longitudinal
linear.  Eq. (\ref{hi-t}) defines the action invariant 
under the gauge transformations (\ref{16}) and
(\ref{17}).  Similarly, eq. (\ref{hi-l}) defines  the action invariant 
under the gauge transformations (\ref{16}) and
(\ref{18}). 

Two formulations of Ref. \cite{KS} 
for the massless multiplet of an integer superspin $s$ 
(with $s=1,2,\ldots$),
longitudinal and transversal, 
contain the following dynamical variables 
respectively:
\bea
\cV^\bot_s &=&
\Big\{H_{\a(s-1)\ad(s-1)}~, ~
\G_{\a(s) \ad(s)}~,
~ \bar{\G}_{\a(s) \ad(s)} \Big\} ~,    \\
\label{12}
\cV^\|_s &=&
\Big\{H_{\a(s-1)\ad(s-1)}~, ~
G_{\a(s) \ad(s)}~,
~ \bar{G}_{\a(s) \ad(s)} \Big\} ~.
\label{13}
\eea
Here $H_{\a(s-1) \ad (s-1)}$ is real, 
$\G_{\a (s) \ad (s)} $ transverse linear and 
$G_{\a (s) \ad (s)}$ 
longitudinal linear tensor superfields.
The case $s=1$
corresponds to the gravitino multiplet 
(see \cite{BK,GGRS} for  reviews).

The gauge transformations  
for the superfields $H_{\a (s-1) \ad (s-1)}$, 
$G_{\a (s) \ad(s)}$ and $\G_{\a(s) \ad (s)}$ 
postulated in \cite{KS} are
\bea
 \d   H_{\a (s-1) \ad (s-1)}&=& \g_{\a (s-1) \ad (s-1)} 
+ {\bar \g}_{\a (s-1) \ad (s-1)}~,
\label{21}\\
 \d  \G_{\a (s) \ad(s)}
 &=& \hf D_{(\a_s} {\bar D}_{(\ad_s} \,
 \g_{\a_1 \dots \a_{s-1} ) \ad_1 \dots \ad_ {s-1})}
-{\rm i}\, s \, \pa_{(\a_s (\ad_s } \,
\g_{\a_1 \dots \a_{s-1} ) \ad_1 \dots \ad_ {s-1})} ~, 
\label{23} \\
\d  G_{\a (s) \ad(s)} &=& 
\hf \bar D_{(\ad_s}  D_{( \a_s} \,
{\bar \g}_{ \a_1 \dots \a_{s-1} ) \ad_1 \dots \ad_ {s-1}) }~,
\label{22}
\eea
with a transverse linear parameter $\g_{\a(s-1) \ad(s-1)}$.  
It can be seen that $\d \G_{\a (s) \ad (s)}$ is transverse
linear.  Eq. (\ref{i-l}) defines the action invariant 
under the gauge transformations (\ref{21}) and
(\ref{22}).  Similarly, eq. (\ref{i-t}) defines  the action invariant 
under the gauge transformations (\ref{21}) and
(\ref{23}).

\sect{Twistor Oscillator Realization}

$~~~~$ In the present  section, we describe a twistor-like oscillator 
realization that is intrinsically related to the superfield structure 
of the dynamical variables  and gauge transformations 
reviewed in the previous section.
This oscillator realization can be used to obtain 
a generating formulation for the massless multiplets
of arbitrary superspin. 

Associated with the left spinor representation 
$(1/2 , 0) $ is a pair of bosonic annihilation $a^\a$ and creation 
$c_\b$ operators,  
\be
[a_\a \,,\, a_\b] = [c_\a \,, \,c_\b] =0~, \qquad
[a^\a \,, \,c_\b ] =\d^\a{}_\b~, \qquad \a,\b=1,2~.
\label{twistor1}
\ee
Similarly, associated with 
 the right spinor representation 
$(0, 1/2 ) $ is a pair  
of bosonic annihilation ${\bar a}^\ad$ and creation 
${\bar c}_\bd$ operators,  
\be
[{\bar a} _\ad \,, \,{\bar a}_\bd ] =[{\bar c} _\ad \,, \,{\bar c}_\bd ] =0~,
\qquad 
[{\bar a} ^\ad \,, \,{\bar c}_\bd ] =
\d^\ad{}_\bd~, \qquad \ad,\bd=1,2~.
\label{twistor}
\ee
The left spinor and the right spinor operators are defined to 
commute with each other.   The  ket $ |\, 0 \rangle $ and bra 
$\langle 0\,| $ vacuum states are defined by 
\be
a_\a \, |\, 0 \rangle = {\bar a}_\ad \, |\, 0 \rangle =0~, 
\qquad \langle 0\,| \, c_\a =  \langle 0\,| \,{\bar  c}_\ad =0 ~,
\qquad \langle 0\, | \,0 \rangle =1~.
\ee
The above commutation relations occur upon  the 
canonical quantization of
a conformally invariant twistor dynamical system, 
see \cite{twistors} and references therein.
It is quite remarkable that the same oscillator 
realization turns out to be dictated by the 
superfield structure of the gauge 
massless higher superspin multiplets.\footnote{In
Vasiliev's  approach to nonlinear higher spin equations of motion, 
one often considers a smaller set 
\newline $~~~~~~$ of oscillators: 
$[y_\a \,, y_\b ] = 2{\rm i}\, \ve_{\a \b} $, 
$[{\bar y}_\ad \,,{\bar y}_\bd ]= 2{\rm i}\, \ve_{\ad \bd}  $, 
$ [{y}_\a \,, {\bar y}_\bd ] =0$, see e.g. 
\cite{Vas2}.}

Along with the annihilation/creation operators introduced,
let us also consider the superspace spinor covariant derivatives, 
\be
\{ D_\a \, , \, D_\b \} = 
\{ {\bar D}_\ad \, , \,{\bar D}_\bd \} =0~, 
\qquad
 \{ D_\a \, , \,{\bar D}_\bd \} = -2{\rm i} \,\pa_{\a \ad}~.
\ee 
We can now define 
the operators 
$\bm{\cal C}  \, \equiv \, c^\a D_\a $ and $\bm{\cal A}
\, \equiv \, a^\a   D_\a$ 
with the following properties
\be
{\bm{\cal C}}^2 = 0 ~, \qquad    {\bm{\cal A} }^2 
= 0~, \qquad 
\{\bm{\cal C} \,, \,   \bm{\cal A}  \} = D^2~.
\label{difgeom1}
\ee
Similar properties hold for the operators
${\Bar {\bm{\cal C}}}  = {\bar c}^\ad {\bar D}_\ad$ and
${\Bar {\bm{\cal A}}}  = {\bar a}^\ad {\bar D}_\ad$, 
\be
{ {\bm{\cal {\Bar C}}}} ^2 = 0 ~, \qquad {\Bar {\bm{\cal A}}}^2 = 0 
~, \qquad 
\{ {\Bar {\bm{\cal C}}}  \,, \, {  \Bar {\bm{\cal A}}  } \} 
= -{\bar D}^2~.
\label{difgeom2}
\ee

Consider a state $|\J_n \rangle  $ in the Fock space of the form 
 \be
|\J_n \rangle  = \J_{\a_1 \cdots \a_n } (z)  \, 
c^{\a_1} \cdots c^{\a_n} |\, 0 \rangle~, 
\qquad 
\J_{(\a_1 \cdots \a_n )}=\J_{\a_1 \cdots \a_n }~.
\ee
Since 
\bea
{\bm {\cal C}} \,|\J_n \rangle &=& D_{(\a_1}  \J_{\a_2 \cdots \a_{n+1}) }
\, c^{\a_1} \cdots c^{a_{n+1}} |\, 0 \rangle~, \non \\
{\bm {\cal A}} \,|\J_n \rangle &=& n\,D^\b  \J_{\b \a_1 \cdots \a_{n-1} }
\, c^{\a_1} \cdots c^{a_{n-1}} |\, 0 \rangle~,
\eea
we obtain 
\bea
{\bm {\cal C}} \,|\J_n \rangle &=& 0 \quad 
\longleftrightarrow \quad 
D_{(\a_1}  \J_{\a_2 \cdots \a_{n+1}) } = 0~, \non \\
{\bm {\cal A}} \,|\J_n \rangle &=& 0 \quad 
\longleftrightarrow \quad 
D^\b  \J_{\b \a_1 \cdots \a_{n-1} }=0~. 
\eea
Now, it is obvious that the transverse and longitudinal 
linear superfields, which were introduced in the previous
section, are intrinsically related to the
operators ${\Bar {\bm {\cal C}}}$ and ${\Bar {\bm {\cal A}}}
$.
Consider states in the Fock space of the form
\bea
|\J_{(k,l)} \rangle  &=& 
\J_{ (\a_1 \cdots \a_k ) (\ad_1 \cdots \ad_l ) }   \, 
c^{\a_1} \cdots c^{\a_k} \,
{\bar c}^{\ad_1} \cdots {\bar c}^{\ad_l} \, 
|\, 0 \rangle~, \non \\
\langle \J_{(k,l)} |  &=& 
\langle 0\,| \, a^{\a_1} \cdots a^{\a_k} \, 
 {\bar a}^{\ad_1} \cdots {\bar a}^{\ad_l} \, 
\J_{ (\a_1 \cdots \a_k ) (\ad_1 \cdots \ad_l ) }   \, 
\label{pfrms}
\eea
Then, the constraint (\ref{transverse}) is equivalent to 
\be
{\Bar {\bm {\cal A}}} \,|\G_{(k,l)} \rangle =0 \quad 
\longleftrightarrow \quad 
\langle \G_{(k,l)} |  \, {\Bar {\bm {\cal C}}} =0~.
\ee
Similarly, the constraint (\ref{longit}) is equivalent to 
\be
{\Bar {\bm {\cal C}}} \, | G_{(k,l)} \rangle =0 \quad 
\longleftrightarrow \quad 
\langle G_{(k,l)} |  \, {\Bar {\bm {\cal A}}} =0~.
\ee

Relations (\ref{difgeom1}) are reminiscent of  famous 
constructions in differential geometry, 
see e.g.  \cite{Goldberg}.
One can consider $\bm \cC$ and $\bm \cA$ to be analogues
of the exterior differential  $d$ and 
the co-differential $\d \propto \ast \, { d} \, \ast$, 
with $\ast$  the Hodge star operation. 
Then, the third relation (\ref{difgeom1}) 
is analogous to the definition of the Laplacian 
$ \{ d\,, \d\} =\D$. Of course, for this analogy 
to be quite solid, 
it would be good to have a `star' operation $\bm \ast$
in superspace\footnote{In 
the work of Ref. \cite{Pform} it was noted there 
is a natural definition for 
the Hodge `star'  operation 
\newline $~~~~~~$ defined on the irreducible 
pre-potentials of lower 
spin 4D, $\cal N$ = 1 gauge theories.}  
with the  properties
\be
 \bm{\cal A} ~\propto~ {\bm \ast} \,{\bm{\cal C}} \, 
{\bm \ast}~, \qquad {\bm \ast} \, {\bm \ast} = {\rm id}.
\ee
Such an operation does exist, and 
it exchanges 
the ket and bra states. 
\be
{\bm \ast}: ~ |\J_{(k,l)} \rangle \quad \longrightarrow 
\quad  \langle \J_{(k,l)} | ~.
\ee
The specific features of the superspace construction, 
as compared with that of differential geometry, 
are given by  the identities
\be 
{\bm \cC} \, {\bm \cA} = -\hf \, {\bm \cN} \, D^2~,
\qquad 
{\Bar {\bm \cC}} \, {\Bar {\bm \cA}} = 
\hf \,{\Bar {\bm \cN}} \, {\bar D}^2~, 
\ee
with $\bm \cN$ and ${\Bar {\bm \cN}} $ the number operators
\bea
{\bm \cN} &=&c_\a \,a^\a~,\qquad 
{\bm \cN}\, |\J_{(k,l)} \rangle =k\,|\J_{(k,l)} \rangle~; \non \\
{\Bar {\bm \cN}}  &=&{\bar c}_\ad \,{\bar a}^\ad~, 
\qquad  {\Bar {\bm \cN}} \, 
|\J_{(k,l)} \rangle =l\,|\J_{(k,l)} \rangle~.
\eea

Now, taking into account the obvious identities
\bea
\{ {\bm {\cal C}}\,, {\Bar {\bm {\cal C}}} \} &=& -2{\rm i} \,c^\a \,{\bar c}^\ad 
\, \pa_{\a \ad} ~, 
\qquad 
\{ {\bm {\cal C}}\,,  {\Bar {\bm {\cal A}}} \} = -2{\rm i} \,c^\a \,{\bar a}^\ad 
\, \pa_{\a \ad}~,
\non \\
\{ {\bm {\cal A}}\,, {\Bar {\bm {\cal C}}} \} &=& -2{\rm i} \,a^\a \,{\bar c}^\ad 
\, \pa_{\a \ad} ~, 
\qquad 
\{ {\bm {\cal A}} \,,   {\Bar {\bm {\cal A}}} \} = -2{\rm i} \,a^\a \,{\bar a}^\ad 
\, \pa_{\a \ad}~,
\label{extalg}
\eea
one can rewrite the higher superspin
gauge transformations 
in terms of Fock space states
and the differential operators
${\bm {\cal C}}$,  ${\bm {\cal A}}$, 
${\Bar {\bm {\cal C}}}$ and ${\Bar {\bm {\cal A}}}$.
In particular, the gauge transformations
 (\ref{16})--(\ref{18}) take the form
\bea
\d | H_{(s,s)} \rangle &=& 
{\Bar {\bm {\cal C}}} \, | \z_{(s,s-1)} \rangle 
- {\bm {\cal C}} \, | {\bar \z}_{(s-1,s)} \rangle ~,
\non \\
\d | \G_{(s-1,s-1)} \rangle &=&
-\hf \, {1 \over s(s+1) } \,
{\Bar {\bm {\cal A}}} \,  {\bm {\cal A}} \, {\bm {\cal C}} \,
| \bar{\z}_{(s-1,s)} \rangle ~, \\
\d | G_{(s-1,s-1)} \rangle &=&
-{1\over s^2(s+1)} \Big( 
 {\bm {\cal A}} \, {\Bar {\bm {\cal A}}}
+(s+1) \,  {\Bar {\bm {\cal A}}}  \,  {\bm {\cal A}}\Big)
{\Bar {\bm {\cal C}}} \, | \z_{(s,s-1)} \rangle  ~,
\non 
\eea
with $\z_{\a(s) \ad(s-1)}$ an unconstrained
spin-tensor.
Similar results follow for  the gauge transformations 
(\ref{21})--(\ref{22}). 
${}$One can express the gauge-invariant
actions (\ref{hi-t})--(\ref{i-t}) in terms 
of the operators  ${\bm {\cal C}}$,  ${\bm {\cal A}}$, 
${\Bar {\bm {\cal C}}}$ and ${\Bar {\bm {\cal A}}}$,
and special Fock space states. In particular, 
one obtains
\bea
D^\a {\bar D}^2 D_\a |\J_{(k,l)} \rangle  &=& 
{1\over k+1} \Big( 
{\bm {\cal C}}\,  \{  {\Bar {\bm {\cal C}}} \,,   {\Bar {\bm {\cal A}}}  \}  
\,{\bm {\cal A}}
-{\bm {\cal A}} \, \{ {\Bar {\bm {\cal C}}} \,,  {\Bar {\bm {\cal A}}}  \}  
\,{\bm {\cal C}} \Big) \, 
|\J_{(k,l)} \rangle \non \\
&=& 
{1\over l+1} \Big( 
{\Bar {\bm {\cal C}}} \,  \{ {\bm {\cal C}}\,, {\bm {\cal A}} \} 
 \,  {\Bar {\bm {\cal A}}}
-   {\Bar {\bm {\cal A}}} \, \{ {\bm {\cal C}}\,, {\bm {\cal A}} \}  
\, {\Bar {\bm {\cal C}}} \Big) \, 
|\J_{(k,l)} \rangle~.
\label{EoM1}
\eea
Thus, it is clear that the Fock space realization discussed in this 
section can be used as an organizing tool for
formulating the dynamics of the higher spin gauge 
superspins. Moreover, this twistor formalism can be used 
 to obtain a generating formulation 
(different from the formulation  developed in \cite{GKS2})
for such supermultiplets. 

Of course, the twistor construction discussed above also appears 
(albeit in a  hidden manner) in the context of more familiar 
4D, $\cN=1$  theories.  
Consider, for instance,   
two Fock space states  
$|V \rangle \equiv |V_{(0,0)} \rangle $ and  
$|\L \rangle \equiv |\L_{(0,0)} \rangle $, 
with $V$ real and $\L$ chiral,  
${\bm {\Bar {\cal C}}}  |\L  \rangle = 0$.  
It is now clear
that 4D, $\cN=1$ Abelian gauge theory, in 
this language, takes the form  
(i)
$ \d \, |V \rangle =   |\L \rangle + |{\bar \L} \rangle  $ for 
the gauge transformation;
(ii)
$|W_{(1,0)} \rangle = -  {\1\over 4} {  \Bar {\bm{\cal A}}  } \,
{\Bar {\bm{\cal C}}}  \,
 {\bm{\cal C}}  |V \rangle $
for the usual field strength; 
and (iii) the result in (\ref{EoM1}) in the
case of $k=l=$ 0 for the equation of motion.  
It is thus clear that the quartet of 
BRST-like operators ${\bm {\cal C}}$,  
${\bm {\cal A}}$, ${\Bar {\bm {\cal C}}}$ and 
${\Bar {\bm {\cal A}}}$
defined in terms of the twistor 
annihilation and creation  operators
can be used to express usual 4D, $\cN=1$ supersymmetric
gauge theories as statements on an associated Fock space.  

It is worth pointing out that 
the operators (\ref{extalg})
have interesting interpretations
when  the states in (\ref{pfrms})
 are associated with  (gauge) differential forms 
(say, a zero-form $| \vf_{(0,0)} \rangle$, 
a one-form $| V_{(1,1)} \rangle$,
two-forms 
$| F_{(2,0)} \rangle$ and $| {\bar F}_{(0,2)} \rangle$)
for which the operators $d$ and $\d$ are defined.
${}$For such states 
\newline $~~~$ (a.)
the operator $\{ {\bm {\cal A}} \,,   
{\Bar {\bm {\cal A}}} \} $
generates the effect of $\d$;  
\newline $~~~$ (b.)
the operators $\{ {\bm {\cal C}} \,,   
{\Bar {\bm {\cal A}}} \} $ and  $\{ {\bm {\cal A}} \,,   
{\Bar {\bm {\cal C}}} \} $ generate 
the (building blocks for)
\newline $~~~~~~~~~$ 
gauge invariant field strengths
and Bianchi identities; and 
\newline $~~~$ 
  (c.) the operator  $\{ 
{\bm {\cal C} } \, , \,
        {\Bar {\bm \cC} }  \}$ generates the
       the gauge transformation of the one-form.
\newline \noindent
So the quartet 
${\bm {\cal C}}$,  ${\bm {\cal A}}$, 
${\Bar {\bm {\cal C}}}$ and ${\Bar {\bm {\cal A}}}$
allows for a factorization of the usual  
 the exterior differential and 
the co-differential  as realized in the Fock space.

As is well-known, the superfield constraints 
in extended super Yang-Mills theories
\cite{GSW} naturally lead to elegant twistorial interpretations
as integrability conditions in spaces with auxiliary dimensions \cite{Witten,GIY,Rosly}, 
and these and related ideas 
apparently culminated in the discovery of the 
profound concept of harmonic superspace \cite{GIKOS}.
Our discussion above demonstrates  that the superfield 
structure of the higher spin gauge supermultiplets
becomes transparent within the twistor 
approach. We believe that this is not accidental, 
and may be of importance in the context of superstring theory.

\sect{Field Strengths and Bianchi Identities:
Half-integer Superspin}

$~~~~$ We now turn to discussing the structure of 
gauge-invariant field strengths and corresponding 
Bianchi identities, which occur in the massless 
higher superspin models.
We first consider the case of half-integer superspin.

In both the transverse and longitudinal formulations, 
there exists a  gauge-invariant field 
chiral strength that is constructed in terms 
of the prepotential $H_{\a(s) \ad(s)} $ only.
It has the form \cite{GGRS}
\be
W_{\a(2s+1)} =
{1 \over 4} {\bar D}^2 \pa _{(\a_1}{}^{\bd_1} 
\cdots \pa _{\a_s}{}^{\bd_s} D_{\a_{2s+1}}
H_{\a_{s+1} \cdots \a_{2s} ) \bd_1\cdots \bd_s } ~, 
\qquad {\bar D}_\bd W_{\a(2s+1)} =0~.
\label{Weyl}
\ee
The other functional-independent strengths 
involve, depending upon the formulation under 
consideration,  the compensators 
$\G_{\a(s-1) \ad(s-1)}$ or 
$G_{\a (s-1) \ad(s-1)}$ 
and/or their conjugates. 
On the mass shell, 
$W_{\a(2s+1)}$ and its conjugate are the only 
non-vanishing gauge-invariant strengths.

\subsection{Transverse Formulation} 
$~~~~$
The equations of motion,  $E^{\bot}{}_{\a(s) \ad(s)}=0 $ 
and $L_{\a(s-1) \ad(s)}=0 $, are given in terms 
of the following gauge invariant field strengths 
\bea
E^{\bot}{}_{\a(s) \ad(s)} &=&{1\over 4} 
D^\b {\bar D}^2 D_\b  H_{\a(s) \ad(s)}
+D_{\a_s}  {\bar D}_{\ad_s}
\G_{\a(s-1) \ad(s-1)}
- {\bar D}_{\ad_s} D_{\a_s} 
{\bar \G}_{\a(s-1) \ad(s-1)} ~, \non \\
L_{\a(s-1) \ad(s) } &=&
-{1\over 4} {\bar D}^2
D^\b  H_{\b \a(s-1)  \ad(s) } \non \\
&& \qquad \quad+ 
{\bar D}_{\ad_s } \Big(
{\bar  \G}_{\a(s-1)  \ad (s-1)} 
+{s+1 \over s} 
\G_{\a(s-1) \ad (s-1)} \Big)~. 
\label{Einst1}
\eea
They can be shown to obey the Bianchi identity
\be
D^\b E^{\bot}{}_{\b \a(s-1) \ad(s)}  =\hf D^2 
{ L}_{\a(s-1) \ad(s) } ~.
\ee
One can also check that 
the fields strengths (\ref{Weyl}) and 
(\ref{Einst1}) are related to each other by 
\bea
D^\b W_{\b \a(2s)} &=& \pa _{(\a_{1}}{}^{\bd_1} 
\cdots \pa _{\a_{s}}{}^{\bd_s} 
\Big\{ E^{\bot}{}_{\a_{s+1} \cdots \a_{2s}) \bd(s)}  \non \\
&&\qquad \quad  +{s \over 2s +1} 
\Big( {\bar D}_{\bd_s} 
{\bar L}_{\a_{s+1} \cdots \a_{2s}) \bd(s-1) } 
- D_{\a_{s+1}} L_{\a_{s+2} \cdots \a_{2s}) \bd(s) }
 \Big) \Big\}~.
\eea
On-shell, this turns into $D^\b W_{\b \a(2s)} =0$. 
As discussed in detail in \cite{BK}, the equations 
${\bar D}_\bd W_{\a(2s+1)} = D^\b W_{\b \a(2s)} =0$
define an irreducible on-shell massless superfield.

\subsection{Longitudinal Formulation}
$~~~~$
The equations of motion,  $E^{\|}{}_{\a(s) \ad(s)}=0 $ 
and $T_{\a(s-2) \ad(s-1)}=0 $, are given in terms 
of the following gauge invariant field strengths 
\bea
 E^{\|}{}_{\a(s) \ad(s)} &=& {1\over 4} 
D^\b {\bar D}^2 D_\b  H_{\a(s) \ad(s)}
-{1 \over 4} \, {s \over 2s +1} 
[D_{\a_s}, {\bar D}_{\ad_s}]\, [D^{\b}, {\bar D}^{\bd}]
H_{\a(s-1) \b \ad(s-1)\bd} \non  \\
&- &s \, \pa_{a_s \ad_s} \, \pa^{\b \bd} 
H_{\a(s-1) \b \ad(s-1)\bd} \non \\
&-&2{\rm i} \, {s \over 2s +1} \, \pa_{\a_s \ad_s} 
\left( G_{\a(s-1)  \ad(s-1)} -{\bar G}_{\a(s-1)  \ad(s-1)} \right)~,
\label{Einst2}\\
T_{\a(s-1) \ad(s-2)}&=& {\bar D}^\bd  \Big({\rm i}s\, 
\pa^{\g \gd} H_{\g \a(s-1) \, \bd \gd \ad(s-2)} 
+{\bar G}_{\a(s-1)  \bd \ad(s-2)} 
-{s+1 \over s} G_{\a(s-1)  \bd \ad(s-2)} \Big)~. 
\non
\eea
The field strengths introduced obey the Bianchi identity
\be 
D^\b  E^{\|}{}_{\b \a(s-1) \ad(s)} = {1 \over 2s+1} 
\Big\{ {\bar D}_{\ad_s} D_{\a_{s-1}}
{\bar T}_{\a(s-2) \ad(s-1)}
-2{\rm i} (s-1) \pa_{\a_{s-1} \ad_s} 
{\bar T}_{\a(s-2) \ad(s-1)} \Big\}~.
\ee

The Bianchi identity relating the strengths (\ref{Weyl}) and 
(\ref{Einst2}) is 
\bea
D^\b W_{\b \a(2s)} &=& \pa _{(\a_{1}}{}^{\bd_1} 
\cdots \pa _{\a_{s}}{}^{\bd_s} 
\, E^{\|}{}_{\a_{s+1} \cdots \a_{2s} )\bd(s)} ~.
\eea

\sect{Field Strengths and Bianchi Identities:
Integer Superspin}
$~~~~$ Here we consider only the longitudinal formulation. 
Let us introduce the completely symmetric tensor \cite{KS}
\be \eqalign{
W_{\a(2s)} = &\hf s \,
 \pa _{(\a_1}{}^{\bd_1} 
\cdots \pa _{\a_{s-1}}{}^{\bd_{s-1}} {\bar D}^{\bd_s}
D_{\a_{s}} G_{\a_{s+1} \cdots \a_{2s} ) \bd_1\cdots \bd_s } \cr
&-{\rm i}\, \pa _{(\a_1}{}^{\bd_1}  \cdots \pa _{\a_{s}}{}^{\bd_{s}} 
G_{\a_{s+1} \cdots \a_{2s} ) \bd_1\cdots \bd_s } 
~. }
\label{Weyl2}
\ee
It can be readily checked that $W_{\a(2s)} $ is gauge invariant, 
and that it is chiral, 
\be 
{\bar D}_\bd W_{\a(2s)} 
=0~.
\ee
If one introduces an unconstrained gauge  prepotentil for 
$G_{\a(s) \ad(s)}$, 
\be 
G_{\a(s) \ad(s)} = {\bar D}_{( \ad_s} \J_{\a(s) \ad_1 \cdots \ad_{s-1})}~,
\label{psi}
\ee 
then the field strength can be expressed in the form \cite{GGRS}
\be
W_{\a(2s)} = {1\over 4} (s+1) {\bar D}^2 
 \pa _{(\a_1}{}^{\bd_1} 
\cdots \pa _{\a_{s-1}}{}^{\bd_{s-1}}  D_{\a_{s}}
\J_{\a_{s+1} \cdots \a_{2s} ) \bd_1\cdots \bd_{s-1} } ~.
\ee

The equations of motion,  $E^{\|}{}_{\a(s-1) \ad(s-1)}=0 $ 
and $T_{\a(s) \ad(s-1)}=0 $, are given in terms 
of the following gauge invariant field strengths 
\be \eqalign{
E^{\|}{}_{\a(s-1) \ad(s-1)} =&{1\over 4} 
D^\b {\bar D}^2 D_\b  H_{\a(s-1) \ad(s-1)} \cr
&+{s\over s+1} \Big(D^{\b}  {\bar D}^{\bd}
G_{\b \a(s-1) \bd \ad(s-1)} - {\bar D}^{\bd} D^{\b} 
{\bar G}_{\b \a(s-1) \bd \ad(s-1)} \Big)~, \cr
T_{\a(s) \ad(s-1)}=& {\bar D}^\bd  \Big(
-\hf {s \over s+1}\, {\bar D}_{(\bd} D_{(\a_s}
H_{\a_1 \cdots \a_{s-1} ) \, \ad_1 \cdots \ad_{s-1})} \cr
&+ {\bar G}_{ \a(s)  \bd  \ad(s-1)} 
+{s \over s +1} G_{\a(s)  \bd \ad(s-1)} \Big)~.  
}
\label{Einst3}
\ee

The Bianchi identities are:
\be \eqalign{
D_{(\a_s} E^{\|}{}_{\a_1 \cdots \a_{s-1}) \bd(s-1)} =& -\hf D^2 
T_{ \a (s) \ad (s-1) }~, \cr
D^\g W_{\g \a(2s-1) } =& -{1\over 8} (s+1) \Big( 
\pa _{(\a_1}{}^{\bd_1} \cdots \pa _{\a_{s-1}}{}^{\bd_{s-1}}  
D^2 T_{\a_s \cdots \a_{2s-1}) \bd (s-1)}  \cr
&-2 \pa _{(\a_1}{}^{\bd_1} \cdots \pa _{\a_{s-1}}{}^{\bd_{s-1}} 
\Big\{ {\bar D}^{\bd_s} D_{\a_s} -2{\rm i} \pa_{\a_s}{}^{\bd_s} \Big\}
{\bar T}_{\a_{s+1} \cdots \a_{2s-1}) \bd (s)} 
~. }
\ee
On-shell, the latter  turns into $D^\b W_{\b \a(2s-1)} =0$, 
and therefore $W_{\a (2s)} $ becomes 
an irreducible on-shell massless superfield \cite{BK}.

\vskip.5cm

\noindent
{\bf Acknowledgements:}\\
SMK is grateful to the Center for String and Particle Theory at the 
University of Maryland for hospitality.
The work of SJG is supported in part  by National Science Foundation
Grant PHY-0099544.
The work of SMK is supported in part by the Australian Research
Council.

\begin{appendix}

\sect{Massless Higher Superspin Actions}

$~~~~$ In this appendix, we collect the gauge-invariant action 
for the massless higher superspin multiplets 
\cite{KSP,KS}.
 
\subsection{Half-integer superspin}

In the transverse formulation, 
the action  reads
\bea
S^{\bot}_{s+1/2}&=&
\Big( - \frac{1}{2}\Big)^s  \int {\rm d}^8z \,
\Big\{ \frac{1}{8} H^{ \a(s) \ad(s) }  D^\b {\bar D}^2 D_\b 
H_{\a(s) \ad(s) }  \non \\
&+& H^{ \a(s) \ad(s) }
\left( D_{\a_s}  {\bar D}_{\ad_s} \G_{\a(s-1) \ad(s-1) }
- {\bar D}_{\ad_s}  D_{\a_s} 
{\bar \G}_{\a (s-1) \ad (s-1) } \right) \non \\
&+&\Big( {\bar \G} \cdot \G
+ \frac{s+1} {s} \, \G \cdot \G ~+~ {\rm c.c.} \Big)
\Big\} ~.
\label{hi-t}
\eea
In the longitudinal formulation, the action is
\bea
S^{\|}_{s+1/2}&=&
\Big(-\hf \Big)^s  \int {\rm d}^8z \, \Big\{
\frac 18 
H^{\a (s) \ad (s) }   D^\b \bar D^2
 D_\b H_{\a (s) \ad (s) }  \non \\
&-& \frac{1}{8} \, \frac{s}{2s+1} \, \Big( \,
\big[ D_{\g}, \bar D_{\gd}\big] H^{\g \a (s-1)\gd  \ad (s-1)}
\, \Big)  \,
\big[ D^\b, \bar D^{\bd}\big]
H_{\b\a(s-1)\bd\ad(s-1)} \,  \non \\
&+& \frac{s}{2}\, \Big( \partial_{\gd} 
H^{\g \a (s-1) \gd \ad (s-1)}  \Big) \,
\partial^{\b\bd}
H_{\b\a(s-1)\bd\ad(s-1)} 
\non \\
&+& 2{\rm i} \, \frac{ s}{2s+1}  \,  \pa_{\g \gd } 
H^{\g \a (s-1) \gd \ad (s-1)}
\Big( G_{\a(s-1) \ad(s-1)} - \bar G_{\a(s-1) \ad(s-1)} \Big)  \non \\
&+& \frac{1}{2s+1} \Big( \bar G \cdot G - \frac{s+1}s G \cdot G
+ {\rm c.c.}\Big)\Big\}   ~.
\label{hi-l}
\eea
The models (\ref{hi-t}) and (\ref{hi-l}) are dually equivalent 
\cite{KSP}.

\subsection{Integer superspin}
In the longitudinal formulation, the action is
\bea
S^{\|}_s &=&
\Big(-\hf  \Big)^s \int {\rm d}^8z \, \Big\{
\frac18 H^{\a(s-1) \ad(s-1)}  
D^\b \bar D^2 D_\b H_{\a(s-1) \ad(s-1) } \non \\
&+& \frac{s}{s+1}  H^{\a(s-1) \ad(s-1)} 
\left( D^\b  \bar D^\bd G_{\b \a (s-1) \bd \ad (s-1)} -
\bar D^{\bd} D^{\b} 
\bar G_{\b\a(s-1) \bd \ad (s-1)} \right) \non \\
&+&\Big( \bar G \cdot G
+  \frac{s}{s+1}\, G \cdot G
+ {\rm c.c.}\Big)\Big\} ~,
\label{i-l}
\eea
while the action in the transversal formulation takes the form
\bea
S^{\bot}_s&=&- \Big(-\hf\Big)^s  \int {\rm d}^8z\, \Big\{-
\frac{1}{8} H^{\a(s-1)\ad(s-1)}  D^\b \bar D^2
 D_\b H_{\a(s-1)\ad(s-1)}  \non \\
&+& \frac{1}{8} \, \frac{s^2}{(s+1)(2s+1)} \,
 \Big(  \big[ D^{\a_s}, \bar D^{\ad_s}\big] 
H^{\a (s-1) \ad(s-1)} \Big) \, 
\big[ D_{(\a_s}, \bar D_{(\ad_s}\big] 
H_{\a_1 \dots \a_{s-1}) \ad_1 \dots \a_{s-1})} \,  \non \\
&+& \hf \, \frac{s^2}{s+1} \,
 \Big( \pa^{\a_s \ad_s}  H^{\a(s-1) \ad (s-1)} \Big)\,
 \pa_{(\a_s (\ad_s}  H_{\a_1 \dots \a_{s-1}) \, 
\ad_1 \dots \ad_{s-1})}   \non \\
&+& 2{\rm i}\, \frac{s}{2s+1} \, H^{\a(s-1) \ad (s-1) }
\pa^{\a_s \ad_s} 
\Big( \G_{ \a(s) \ad(s) } - \bar\G_{\a(s) \ad(s)} \Big) \non \\
&+& \frac{1}{2s+1} \, \Big( \bar\G \cdot \G
-  \frac{s+1}{s} \, \G \cdot \G + {\rm c.c.} \Big) \Big\}~.
\label{i-t}
\eea
The models (\ref{i-l}) and (\ref{i-t}) are dually equivalent 
\cite{KS}.

\sect{Dual Formulations for the Massive Vector \newline Multiplet}

$~~~~$ This appendix contains three dually equivalent 
realizations for the massive  gravitino multiplet.
Only the first realization possesses the property that 
the compensating multiplet can be completely gauged 
away.

Consider the St\"uckelberg formulation for the massive vector 
multiplet
\bea
\label{chiral}
S_{\rm I} = {1\over 4}  \int {\rm d}^6z \,
W^\a W_\a &+& \hf m^2 \int {\rm d}^8z \, V^2 \non \\
+ \int {\rm d}^8z \, {\bar \F} \F 
&-& m  \int{\rm d}^8z \, V\,({\bar \F} + \F)~,
\eea
with
$W_\a = -{ 1 \over 4} {\bar D}^2 D_\a V$,
and $\F$ a chiral superfield,
${\bar D}_\ad \F =0$. The action is invariant under 
the gauge transformations 
\be
\d V = \L +{\bar \L}~, \qquad \d \F = m\, \L ~, 
\qquad \quad {\bar D}_\ad \L =0~.
\ee
This gauge freedom can be used to gauge away $\F$.

The model (\ref{chiral}) possesses a dual formulation in which 
the chiral multiplet is replaced by the tensor multiplet
described by a real linear  superfield
\be
L = \hf( D^\a \eta_\a + {\bar D}_\ad {\bar \eta}^\ad)~, 
\qquad  \quad {\bar D}_\ad \eta_\a =0~,
\label{ten-field-strength}
\ee
with $\eta_\a$ an unconstrained chiral spinor superfield.
The dual action\footnote{See \cite{K} for $\cN=2$ 
supersymmetric generalizations of the models
considered here.}   
is \cite{Siegel}
\bea
S_{\rm II} &=&  {1\over 4}  \int {\rm d}^6z \,W^\a W_\a
- \hf \intss\, L^2 
+m \intss\, L \,V   
\non \\
&=&  {1\over 4}  \int {\rm d}^6z \,W^\a W_\a
- \hf \intss\, L^2 
- \hf m \left\{  \int {\rm d}^6z \, W^\a \eta_\a
+{\rm c.c.} \right\} ~.
\label{tensor}
\eea
This action remains invariant under the following gauge transformations
\bea
\d V &=& \L +{\bar \L}~, \quad \quad  
 {\bar D}_\ad \L =0  \non \\
\d \eta_\a &=& {\rm i} \,{\bar D}^2 D_\a K~, \quad \quad {\bar K } = K
\eea
which are characteristic of  the massless vector multiplet and the massless
tensor multiplet respectively. 
Unlike the realization (\ref{chiral}), the chiral spinor compensator $\eta_\a$ 
cannot be gauged away even on the mass shell.
 
The model (\ref{chiral}) possesses another  dual formulation in which 
the chiral  multiplet is replaced by the so-called nonminimal scalar multiplet
described by a complex linear superfield
\be
\G = {\bar D}_\ad {\bar \U}^\ad ~, 
\ee
with $\U_\a$ an unconstrained  spinor superfield.
\bea
\label{nonminimal}
S_{\rm III} = {1\over 4}  \int {\rm d}^6z \,
W^\a W_\a &-& \hf m^2 \int {\rm d}^8z \, V^2 \non \\
- \int {\rm d}^8z \, {\bar \G} \G 
&+& m  \int{\rm d}^8z \, V\,({\bar \G} + \G)~.
\eea
The corresponding gauge invariance is
\be
\d V = \L +{\bar \L}~, \qquad \d \G = m\, \L ~, 
\qquad \quad {\bar D}_\ad \L =0~.
\ee
The compensator $\G$ cannot be gauged away even 
on the mass shell.

The fate of the mass term  $m^2 \int{\rm d}^8z\,V^2$ 
in the models (\ref{chiral}), (\ref{tensor}) and (\ref{nonminimal})
is clearly very distinctive.

\sect{Massive Gravitino Multiplet as a Gauge Theory}

$~~~~$ As discussed in the introduction,  
there are reasons to expect that 
a massive superspin-$s$ multiplet can  be described as 
a gauge-invariant dynamical system involving massless 
multiplets of superspins $s,s-1/2, \ldots ,0$.  
To our knowledge, 
no prior realization of this approach has been presented
in 4D, $\cal N$ = 1 superspace for a massive superspin-$s$ 
multiplet for $s >3/2$. 

In this appendix, we present the simplest nontrivial 
case -- massive gravitino mutiplet corresponding to $s=1$.
These results will show that the program described in
in the introduction
works in this  specific case,  
and thus it is encouraging for the future pursuit 
of such realizations of this
approach for all higher superspin cases.

In the case $s=1$, the integer-superspin 
longitudinal formulation\footnote{At $s=1$, 
the  integer-superspin 
transverse model  (\ref{i-t}) 
provides a non-minimal
off-shell realization   \newline $~~~~~~$ for the massless 
gravitino multiplet \cite{KSP}  formulated 
in terms of an unconstrained real scalar
$H$ and  \newline $~~~~~~$  Majorana 
$\g$-traceless
spin-vector ${\bf \J}_a=
(\J_{a \b}, {\bar \J}_{a}{}^\bd)$,
with $\g^a {\bf  \J}_a=0$.}
(\ref{i-l}) 
describes the off-shell massless gravitino multiplet
introduced first in  \cite{FV,deWvH}
at the component level and then formulated
in \cite{GS} in terms of superfields.
The dynamical variables are (i) a real scalar superfield $H$
and (ii) an unconstrained  spinor superfield $\J_\a$ 
that generates $G_{\a \ad}$ according to (\ref{psi}).
The action can be represented in the form 
\be \eqalign{
S^{\|}_{(1,3/2)}[\J , H]
\,=\, &\hat{S} [\J ]
- {1\over 4}
 \int {\rm d}^8z\,\Big\{ 
\J^\a {\bar D}^2 D_\a H 
+{\bar \J}_\ad D^2 {\bar D}^\ad H  
\Big\}~, \cr
&-{1\over 16}   \int {\rm d}^8z\,
 H D^\a {\bar D}^2 D_\a \, H 
 ~,
}
\label{tino}
\ee
where
\bea
  \hat{S}[\J ]
&=& \int {\rm d}^8z\, \Big\{
D^\a{\bar \J}^{\dot\a}{\bar D}_{\dot\a}\J_\a
-\frac 14{\bar D}^{\dot\a}\J^\a{\bar D}_{\dot\a}\J_\a
-\frac 14 D_\a {\bar \J}_{\dot\a} D^\a {\bar \J}^{\dot\a}
\Big\}~.
\label{S-hat}
\eea
In accordance with (\ref{23}), the gauge freedom is:
\bea
\d H = D^\b L_\b + {\bar D}_\bd {\bar L}^\bd~, 
\quad 
\d \J_\a = \eta_\a  +\hf D_\a D^\b L_\b ~, 
\qquad {\bar D}_\ad \eta_\a =0~,
\eea
with the gauge parameter $L_\a$ being an unconstrained
spinor. It is obvious that 
$H$ can be completely gauged away.

The massive extension of (\ref{tino}) 
was obtained in \cite{Ph,BGKP}
\bea
S[ \J,H] &=&
S^{\|}_{(1,3/2)}[\J , H]
+m \int {\rm d}^8z\,\Big\{
\J^2
+{\bar \J}^2
-{1\over 4} m \,H^2
+ \hf H\Big( D  \J
+ {\bar D}  {\bar \J}  \Big)
\Big\}~. 
\label{ss1final}
\eea
In this model for massive gravitino multilet,
it turns out that   
the degrees of freedom associated with $H$ 
can be integrated out.  
Indeed, let us implement the following transformation
\be 
\J_\a ~\to ~ \tilde{\J}_\a = \J_\a 
+{1\over 16m} \,{\bar D}^2 D_\a \,H
\label{shift1}
\ee
in the action. This leads to
\be \eqalign{
S[ \tilde{\J},H] =&
\hat{S}[\J ] -{1\over 4} m^2  \int {\rm d}^8z\,\Big\{
H^2 - {2\over m} H \Big(
D^\a (\J_\a + {1 \over 4m}\,{\bar D}^2 \,\J_\a )
~+~{\rm c.c.} \Big) \Big\}\cr
&+m \int {\rm d}^8z\,\Big\{
\J^2 +{\bar \J}^2 \Big\} 
~.
} \ee
As is seen, the superfield $H$ has become auxiliary, 
and therefore it can be eliminated. 
In conjunction with the  shift
\be 
\J_\a ~\to ~ 
\J_\a 
-{1\over 4m} \,{\bar D}^2 \,\J_\a~,
\ee
one then ends up with 
\be \eqalign{
S[\J] = &\hat{S}[\J ]
+ {1\over 4} \int {\rm d}^8z\,
\Big( D\,\J +
{\bar D} \,{\bar \J}  \Big)^2
-{1\over 2}  \int {\rm d}^8z\,\
\Big\{ \J^\a {\bar D}^2 \J_\a
+ {\bar \J}_\ad D^2 {\bar \J}^\ad \Big\}
 \cr
&+ m \int {\rm d}^8z\,
\Big\{\J^2 + {\bar \J}^2\Big\}~, }
\label{mgravitino2}
\ee
formulated solely in terms of $\J_\a$
and its conjugate.
This action was obtained in \cite{BGKP}
by applying a duality transformation to 
$S[ \J,H]$ (this duality transformation converts $H$ into 
a chiral spinor superfield, and the latter is then integrated
out).
Upon the  
trivial rescaling 
 \be
\J_\a ~\to ~{\rm i}  \,\J_\a~, 
\qquad
m ~\to ~ -m ~,
\ee
the action (\ref{mgravitino2}) takes the form
\bea
S_m[\J] = \hat{S}[\J ]
&-&{1\over 4} \int {\rm d}^8z\,
\Big( D\J -
{\bar D} {\bar \J}  \Big)^2
+ m \int {\rm d}^8z\,
\Big\{\J^2 + {\bar \J}^2\Big\}~.
\label{mgravitino3}
\eea
${}$Finally, the transformation 
\be
\J_\a ~\to ~{\rm e}^{{\rm i} \,\p/4} \,\J_\a~, 
\qquad
D_\a ~\to ~{\rm e}^{ {\rm i} \,\p /4} \,D_\a
\label{equivalence}
\ee
turns (\ref{mgravitino3}) into\footnote{In our previous 
publication \cite{BGKP}, 
the models (\ref{mgravitino3}) 
and (\ref{mgravitino4}) 
were mistakenly treated as  \newline $~~~~~~$ 
different, albeit similar, 
off-shell realizations for the massive gravitino multiplet.
The trans-  \newline $~~~~~~$ formation (\ref{equivalence}) 
establishes 
the equivalence of the two realizations.}  
\bea
S'_m [\J]
=  \hat{S}[\J ]
+ {1\over 4}  \int {\rm d}^8z\,
\Big(  D\J + {\bar D} {\bar \J}
\Big)^2
+ {\rm i} \, m \int {\rm d}^8z\,
\Big\{  \J^2 - {\bar \J}^2
\Big\}~.
\label{mgravitino4}
\eea
The action for massive gravitino multiplet
(\ref{mgravitino3}), 
or its equivalent form (\ref{mgravitino4}), 
was discovered many years ago  
\cite{OS2}. Its massless counterpart, 
\bea 
S_{\rm OS}[\J] = \hat{S}[\J ]
&-&{1\over 4} \int {\rm d}^8z\,
\Big( D\J -
{\bar D} {\bar \J}  \Big)^2~,
\label{OS}
\eea
is the Ogievetsky-Sokatchev  action 
for massless gravitino multiplet \cite{OS}.
The massless actions (\ref{tino}) and (\ref{OS}) 
are related to each other by a duality 
transformation, see  \cite{BK} for a review.
As demonstrated in \cite{GS}, 
the actions  $S_{\rm OS}[\J] $ and 
$\hat{S}[\J ]$ (the latter being a gauge fixed version 
of (\ref{tino})) are the only possible off-shell realizations 
for massless gravitino multiplet in terms  of a single 
spinor superfield.

By analogy with \cite{GGRS}, 
let us apply a transformation 
\be 
\J_\a ~\to ~  \check{\J}_\a = \J_\a +D_\a \,U 
+ {\rm i} \, \S_\a~,  \qquad U\neq {\bar U} ~, 
\qquad {\bar D}_\ad \, \S_\a = 0
\label{nicered}
\ee
to the action $S_{\rm OS}[\J] $. One obtains
\be \eqalign{ 
S_{\rm OS}[\check{\J}] = &S_{\rm OS}[\J] 
-{1\over 4}   \int {\rm d}^8z\,
\Big\{ (U+ {\bar U}) \, 
D^\a {\bar D}^2 D_\a  (U+ {\bar U}) 
\, -\, (D \S +{\bar D} {\bar \S})^2 \Big\}
\cr
&+ \hf  \int {\rm d}^8z\,\Big\{ 
\J^\a {\bar D}^2 D_\a 
(U+ {\bar U}) 
+{\rm i} \,
\J^\a D_\a 
(D \S  + {\bar D} {\bar \S})  
~+~ {\rm c.c.} \Big\} ~.  }
\ee
${}$From here it is obvious that 
the Ogievetsky-Sokatchev  action (\ref{OS}) 
possesses the following gauge freedom
\be
\d \J_\a = {\rm i} \, D_\a K_1 +{\rm i} \,
{\bar D}^2 D_\a K_2~, 
\qquad 
{\bar K}_i = K_i~.
\ee
Now, following \cite{AB}, 
in (\ref{nicered}) we choose
\bea
U=\hf V  -{ {\rm i} \over 4m} \,  
(D \eta  + {\bar D} {\bar \eta}) 
~, \qquad 
\S_\a = \eta_\a + 
{{\rm i}  \over 4m} \,
{\bar D}^2 D_\a \,V~, 
\eea
with $V$  a real scalar, 
and $\eta_\a$ a chiral spinor. 
With this choice, 
the massive action (\ref{mgravitino3})
turns \cite{AB} into 
\be \eqalign{
S_m[\check{\J}] =  &S_{\rm OS}[\J ]
+{1\over 4}   \int {\rm d}^8z\,
\Big\{ V D^\a {\bar D}^2 D_\a \, V 
- (D \eta +{\bar D} {\bar \eta})^2 \Big\} 
\cr
&+ m \int {\rm d}^8z\,
\Big\{ ( \J  +\hf DV +{\rm i} \,\eta )^2 ~+~{\rm c.c.}
\Big\}~.  }
\label{mgravitino5}
\ee
In the massless limit, $m\to 0$, 
this action becomes a sum
(with the correct signs!)  of the
gauge massless actions for (i) gravitino multilet; 
(ii) vector multiplet; (iii) tensor multiplet.

In conclusion, it is worth pointing  
out one final possible 
implication of  this discussion.  
Many years ago \cite{TEXAS}, one of the authors (SJG) 
conjectured that in string field theory, there might exist a limit 
in which all  higher spin states become massless 
and yet the theory retains its unitary character.  
Within string theory, all masses are proportional to the
string tension.  In the limit of no tension, all masses approach zero.
Although the present example is far away from being a proof
of this, the extension of the present example along the lines 
described by the works in \cite{KZ,Zinoviev} 
is consistent with this conjecture.

\end{appendix}

\small{

}

\end{document}